\documentstyle[twocolumn,aps,epsfig]{revtex}
\begin{document}
\newcommand{\beq}{\begin{equation}}
\newcommand{\eeq}{\end{equation}}
\newcommand{\bea}{\begin{eqnarray}}
\newcommand{\eea}{\end{eqnarray}}
\newcommand{\bfig}{\begin{figure}}
\newcommand{\efig}{\end{figure}}
\newcommand{\ie}{{\it i.e.}}
\newcommand{\bce}{\begin{center}}
\newcommand{\ece}{\end{center}}
\newcommand{\eg}{{\it e.g.}}
\newcommand{\etal}{{\it et al.}}
\def\lsim{\mathrel{\rlap{\lower4pt\hbox{\hskip1pt$\sim$}}
    \raise1pt\hbox{$<$}}}         
\def\gsim{\mathrel{\rlap{\lower4pt\hbox{\hskip1pt$\sim$}}
    \raise1pt\hbox{$>$}}}         
\twocolumn[\hsize\textwidth\columnwidth\hsize\csname @twocolumnfalse\endcsname
\title
{Low-Mass Dileptons at the CERN-SpS: Evidence for Chiral Restoration?}
 
\author
{Ralf Rapp$^1$ and Jochen Wambach$^2$}
 
\address
{1) Department of Physics and Astronomy, State University of New York, 
    Stony Brook, NY 11794-3800, U.S.A. \\
 2) Institut f\"ur Kernphysik, TU Darmstadt, Schlo{\ss}gartenstr.~9,
    D-64289 Darmstadt, Germany}
 
\maketitle
 
\begin{abstract}
Using a rather complete description of the in-medium 
$\rho$ spectral function - being constrained by various independent
experimental information - we calculate pertinent 
dilepton production rates from hot and dense hadronic matter. 
The strong broadening of the $\rho$ resonance  entails a 
reminiscence to perturbative $q\bar q$ annihilation rates 
in the vicinity of the phase boundary.
The application to dilepton observables
in Pb(158AGeV)+Au collisions - incorporating recent information on the  
hadro-chemical  composition at CERN-SpS energies - essentially supports 
the broadening scenario. Possible implications for the nature of 
chiral symmetry restoration are outlined. 
\end{abstract}

]
 
\section{Introduction}
In the low-energy sector ($q^2\le 1-2 $~GeV$^2$) the properties of 
Quantum Chromodynamics are governed by (approximate) 
chiral symmetry and its dynamical breaking. In hot and dense matter
as created through energetic collisions of heavy nuclei chiral symmetry
is believed to be restored, associated with the disappearance of the 
chiral condensate and a substantial reshaping of the low-lying 
hadronic spectrum. As the chiral condensate does not represent a 
physical observable, the detection of signatures for chiral restoration 
has to focus on  medium effects in hadronic properties.  
Dileptons as penetrating probes provide the opportunity   
to study rather directly the in-medium modifications of 
light vector mesons through their decays $V\to l^+l^-$. 

Measurements of low-mass dilepton spectra in heavy-ion collisions at 
CERN-SpS energies~\cite{CERES,HELIOS3} have revealed appreciable 
radiation beyond final-state hadron decays. 
However, the spectral shape of this additional contribution - stemming
from the interaction phase of the hadronic fireball - is not easily 
accounted for. In particular, using the vacuum vector meson line shapes, 
the enhancement around  $M_{ll}\simeq (0.2-0.6)$~GeV remains unexplained. 
At the same time the yield in the $\rho$/$\omega$ mass region tends 
to be overestimated. The central question then is whether one has 
possibly observed signatures of (the onset of) chiral symmetry restoration, 
and, if so, {\em how} it manifests itself in the vector channel.   
 
Several theoretical approaches have been pursued to study  
medium effects in dilepton production rates from hot/dense matter, 
\eg,  the 'dropping rho mass' scenario (within a mean-field type 
treatment)~\cite{BR91,CEK,LKB}, the chiral reduction formalism (based on 
chiral Ward identities in connection with low-density 
expansions)~\cite{SYZ}, chiral Lagrangian frameworks~\cite{Song}
 or many-body-type calculations of in-medium
vector-meson spectral functions~\cite{RCW,FrPi,KKW97,PPLLM}. 
As a further complication, the quantitative impact of in-medium 
effects on the final spectra depends on another important ingredient, 
namely the modeling of the space-time evolution of the global heavy-ion 
collision dynamics. Here, transport calculations seem to give
substantially larger dilepton yields than hydrodynamic
simula\-tions. A reason for this discrepancy might be related to an 
overpopulation of pion phase space (enhancing the $\pi\pi\to\rho\to ll$
annihilation contribution).

In the present article we will focus on the in-medium spectral
function approach for the elementary dilepton production processes, 
presenting its most recent developments, thereby noting 
a surprising reminiscence to (perturbative) quark-based calculations. 
We then apply the medium-modified dilepton rates to the calculation of 
spectra from $\pi\pi\to\rho\to ee$ annihilation in 
30\% central Pb(158~AGeV)+Au collisions
and perform a detailed comparison to various projections of the recent
data from the CERES/NA45 collaboration. Special care is taken to adopt
a realistic temperature/density profile of the collision dynamics 
using a thermal fireball expansion with a particle composition 
consistent with recent hadro-chemical analyses at 
SpS energies~\cite{pbm98,CR99}. The subsequent evolution
towards thermal freezeout is determined by assuming entropy as 
well as pion- and kaon-number conservation which induces 
the build-up of finite pion-/kaon-chemical potentials~\cite{Bebie}. 
 
\section{Production Rates in Hot/Dense Matter} 

The radiation of dileptons from a hot and dense
medium characterized by a temperature $T$ and baryon chemical potential 
$\mu_B$,
\beq
\frac{d^8N_{l^+l^-}}{d^4x d^4q} = -\frac{\alpha^2}{\pi^3 M^2} \ f^B(q_0;T) \
{\rm Im} \Pi_{\rm em}(q_0,\vec q;\mu_B,T) \ , 
\eeq
is governed by the thermal expectation value of the (retarded)  
electromagnetic current-current correlator, $\Pi_{\rm em}$ 
($f^B$: Bose function). The latter can 
be written in terms of its longitudinal and transverse projections as  
\beq
\Pi_{\rm em}(q_0,\vec q)=\frac{1}{3}
\left[\Pi^L_{\rm em}(q_0,\vec q) +
2\Pi^T_{\rm em}(q_0,\vec q) \right] \ ,
\eeq
both depending separately on energy and momentum.   
Below momentum transfers of about 1~GeV, the hadronic part of the e.m.
current can be accurately  saturated by the light vector mesons 
within the well-established vector dominance model (VDM). The correlator is 
then expressed through the imaginary parts of the vector meson
propagators (= spectral functions) as
\beq
{\rm Im} \Pi_{\rm em} = \sum\limits_{V=\rho,\omega,\phi} 
\frac{(m_V^{(0)})^4}{g_V^2} {\rm Im} D_V \ . 
\eeq
In the following we will neglect the small contributions from the 
isoscalar part and concentrate on the $\rho$ meson which plays by far
the dominant role for the time scales involved in heavy-ion reactions. 

Along the lines of our earlier analyses~\cite{RCW} the $\rho$ 
propagator in hot hadronic matter, 
\beq
D_\rho^{L,T}=\frac{1}{M^2-(m_\rho^{(0)})^2-\Sigma_{\rho\pi\pi}^{L,T}
-\Sigma_{\rho M}^{L,T} -\Sigma_{\rho B}^{L,T} } \ ,  
\label{Drho}
\eeq
is evaluated in terms of various contributions entering 
its in-medium selfenergy $\Sigma_\rho$. The meson gas effects (encoded in
$\Sigma_{\rho M}$) are accounted for following Ref.~\cite{RG99}
through interactions with the most abundant thermal 
$\pi$, $K$ and $\rho$ mesons saturated with a rather complete set of
$s$-channel resonances $R$ up to 1.3~GeV.  The interaction vertices have
been constrained by both hadronic ($R\to P\rho$) {\em and} 
radiative ($R\to P\gamma$) decay branchings. Also, the Bose enhancement
for the in-medium $\rho\pi\pi$ width has been included in 
$\Sigma_{\rho\pi\pi}$. 

The $\rho$ modifications in nuclear matter are based on the model
constructed in Refs.~\cite{RCW,RUBW,UBRW}, where direct $\rho N\to B$ 
interactions ($B$=$N$, $\Delta$, $N$(1520), $\Delta$(1700), 
$N$(1720), $\dots$, encoded in $\Sigma_{\rho B}$)
as well as pion cloud modifications through $\pi NN^{-1}$ and 
$ \pi\Delta N^{-1}$ excitations (entering $\Sigma_{\rho\pi\pi}$) 
were calculated. Again, the model parameters have been
thoroughly constrained by analyzing photoabsorption spectra
on nucleons and nuclei as well as $\pi N \to \rho N $ scattering data.
The latter enforce a rather soft form factor on the $\pi NN$
vertex. 
Also, since the VDM is less accurate in the baryonic sector, an improved
VDM coupling~\cite{KLZ} has been employed for the (transverse)
$\gamma N \to B$ transition form factors so that 
\beq
{\rm Im } \Pi_{\rm em} = \frac{1}{3 g_{\rho\pi\pi}^2} \left[
{\cal F}^L + 2 {\cal F}^T \right]  
\eeq 
with 
\bea
{\cal F}^L &=& -(m_\rho^{(0)})^4 \ {\rm Im} D_\rho^L 
\nonumber\\ 
{\cal F}^T &=& 
-{\rm Im}[\Sigma^T_{\rho\pi\pi}+\Sigma_{\rho M}^T] |d_\rho-1|^2
  -{\rm Im}\Sigma^T_{\rho B} |d_\rho-r_B|^2
\nonumber\\
d_\rho &=&
\frac{ M^2-\Sigma^T_{\rho\pi\pi}-\Sigma_{\rho M}^T  - r_B \Sigma^T_{\rho B} }
{ M^2-(m_\rho^{(0)})^2-\Sigma^T_{\rho\pi\pi}-\Sigma_{\rho M}^T
-\Sigma^T_{\rho B} } \ ,  
\eea
cf.~Refs.~\cite{FrPi,RUBW} (here $r_B=0.7$, the ratio between the actual
$\gamma NB$ coupling and its value in the simple VDM).  

\bfig
\epsfig{file=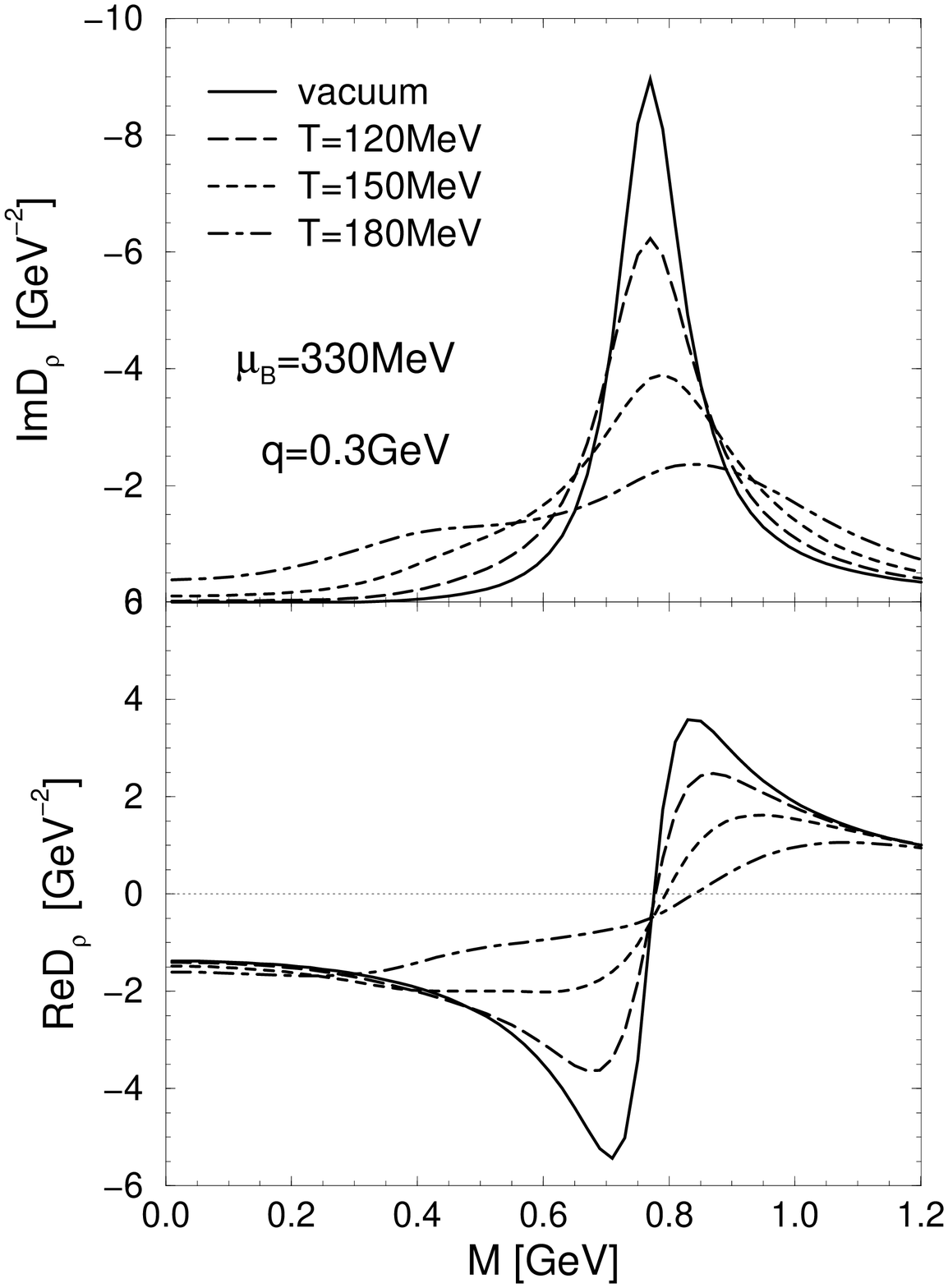,width=8cm}
\caption{Real (lower panel) and imaginary (upper panel) part of the
$\rho$ propagator in hot and dense hadronic matter at fixed
baryon chemical potential $\mu_B=330$~MeV and temperatures $T=180$~MeV
(corresponding to nucleon and total baryon densities of
$\varrho_N=0.67\varrho_0$ and $\varrho_B=2.6\varrho_0$),
$T=150$~MeV ($\varrho_N=0.25\varrho_0$, $\varrho_B=0.72\varrho_0$) and
$T=120$~MeV ($\rho_N=0.06\rho_0$, $\rho_B=0.13\rho_0$).}
\label{fig_drho}
\efig

At finite temperatures two additional features appear in the baryonic
sector. Firstly, the Fermi-Dirac distribution functions are replaced by 
thermal ones~\cite{RW94,RCW}, which, at the temperatures of interest here 
($T\simeq 150$~MeV), are substantially smeared as compared to the 
zero-temperature $\Theta$-functions. Secondly, a large fraction
of the nucleons is excited into baryonic resonances which, in turn, 
are no longer active in nucleon-driven effects. However, as 
conjectured in Ref.~\cite{BLRRW}, a baryonic resonance $B_1$ might still have 
excitations of type $\rho B_2 B_1^{-1}$ built on it. 
The particle data table~\cite{PDG96} indeed supports this hypothesis: \eg,
the $\Sigma(1670)$ (which is a well-established four-star resonance
with spin-isospin $IJ^P=1 \frac{3}{2}^-$), when interpreted  as a $\rho \Sigma$
(or $\rho\Lambda$) 'resonance',   very much resembles the quantum
numbers and excitation energy  ($\Delta E\simeq 500-700$~MeV)
of the $\rho N\to N(1520)$ transition. In addition,
the branching ratio  of $\Sigma(1670)$ decays into 'simple' final states
such as $N\bar K, \Sigma \pi$ or $\Lambda\pi$ is significantly less than 
100\% .
Similar excitations on non-strange baryonic resonances are more
difficult to identify as the latter themselves decay strongly via
pion emission (\ie, the $B_2\to B_1 \rho$ decay is immediately followed
by further $B_2\to N\pi$ and $\rho\to \pi\pi$ decays). Nevertheless,
from pure quantum numbers it is tempting to associate, \eg,
$\Delta(1930)\Delta^{-1}$ or $N(2080)N(1440)^{-1}$ excitations
with $S$-wave  'Rhosobar' states. We have conservatively estimated 
their coupling constants
to give reasonable branching fractions into $\rho B_1$ and $\gamma B_1$
(the latter are typically in the range of 0.2--0.7~MeV). Due
to the large total widths involved, their impact on the $\rho$ propagator
is weaker as compared to excitations on nucleons at equal densities;
\eg, at T=150~MeV and $\mu_B=436$~MeV, where, in chemical equilibrium,
 about 2/3 of the baryons are in excited states, the additional broadening
from the calculated $B_2B_1^{-1}$ excitations as compared to 
$BN^{-1}$ ones amounts to 
about 40\%. Concerning the pion cloud modifications, we do not
explicitly compute the 'Pisobars' on excited resonances but approximate
their effect by using an effective nucleon  density 
$\varrho_{eff}= \varrho_N+0.5\varrho_{B^*}$.

The final result for the $\rho$ spectral function in hot hadronic matter
is displayed in Fig.~\ref{fig_drho} at fixed three-momentum and
chemical potentials $\mu_\pi=0$, $\mu_B=330$~MeV for pions and baryons, 
respectively. One observes a strong broadening
with increasing temperature and density. Comparing to the pure meson
gas results of Ref.~\cite{RG99} (cf.~Fig.~3 therein), it becomes clear that 
especially the low-mass enhancement around $M\simeq 0.4$~GeV is 
largely driven by baryons\footnote{With respect to the conditions 
realized in URHIC's at the full SpS energy (160-200~AGeV) this statement 
has to be taken with care as finite pion chemical potentials might
arise towards freezeout which enhances the contributions from the 
meson gas.}.   

The pertinent three-momentum integrated dilepton production rates 
are shown in Fig.~\ref{fig_dlrates}. 
At moderate temperature and density (upper panel), one still recognizes a  
remnant of the $\rho$ peak, which, however, is entirely wiped out 
under conditions expected to be close to the phase boundary (lower panel). 
This  provokes a comparison with quark-gluon based rate calculations,
which, for simplicity, have been evaluated in terms of lowest order 
$O(\alpha\alpha_s^0$) $q\bar q$ annihilation~\cite{CFR},
\bea
\frac{dR_{q\bar q \to ee}}{d^4q} &=& \frac{\alpha^2}{4\pi^4}
\frac{T}{q} f^B(q_0;T) \sum\limits_q e_q^2 
\nonumber\\
&\times& \ln
\frac{\left(x_-+y\right) \left( x_++\exp[-\mu_q/T]\right)}
{\left(x_++y\right) \left( x_-+\exp[-\mu_q/T]\right)}
\label{qqrate0}
\eea
with $x_\pm=\exp[-(q_0\pm q)/2T]$, $y=\exp[-(q_0+\mu_q)/T]$.
In the vacuum the $q\bar q$ rates are known 
to coincide with the hadronic description for invariant 
masses $M\ge1.5$~GeV as marked by the cross section ratio 
$\sigma(e^+e^-\to hadrons)/\sigma(e^+e^-\to\mu^+\mu^-)$ for the inverse 
process of $e^+e^-$ annihilation (up to additional resonance structures in the 
vicinity of heavy-quark thresholds such as $c\bar c$).   
It seems that the in-medium hadronic rates indeed approach the partonic
ones rather quickly leading to an approximate  
agreement at the highest temperatures/densities 
for masses of about 0.5--1~GeV (the deviation at masses $M>1$~GeV is due 
to the incomplete description of the vector correlator in vacuum which does
not include more than two-pion states; the discrepancy at low masses 
might be reduced once higher order $\alpha_S$ corrections are included, in 
particular soft Bremsstrahlung-type processes). 
A tempting interpretation of this behavior is a lowering of the 
'quark-hadron duality threshold' in hot and dense hadronic matter.    
We will return to this issue in the discussion in Sect.~\ref{sec_disc}. 

\bfig[htb]
\bce
\epsfig{figure=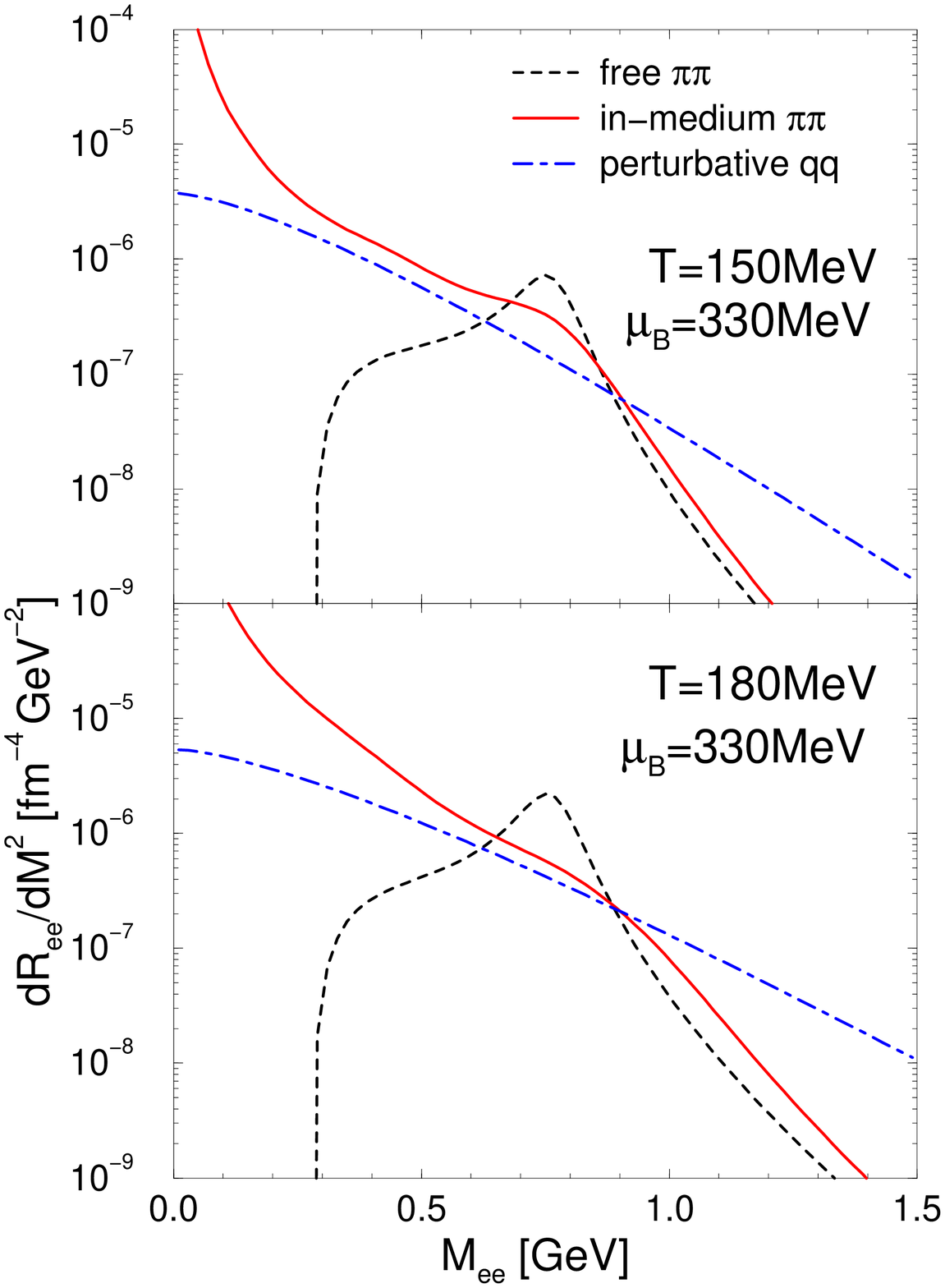,width=8cm}
\ece
\caption{Comparison of hadronic and partonic dilepton production rates 
(integrated over 3-momentum) in hot and dense matter at fixed baryon
chemical potential (and equivalent quark chemical potential 
$\mu_q$=$\mu_B/3$) for  T=150~MeV (upper panel) and T=180~MeV (lower panel);
dashed and full lines: $\pi\pi$ annihilation using the free
and the in-medium $\rho$ spectral function, respectively; dashed-dotted
lines: lowest-order quark-antiquark annihilation.}
\label{fig_dlrates}
\efig

\section{Low-Mass Dilepton Observables in Pb(158~AGeV)+Au }
For a sensible application of the in-medium vector correlator to
calculate dilepton production in URHIC's a realistic temperature
and density profile is required. Microscopic
transport or hydrodynamical simulations, \eg,  have proven very 
successful in describing the final hadron observables. 
However, the associated  dilepton yields, based on
identical production rates, may differ appreciably~\cite{R99}, cf., \eg, 
Refs.~\cite{CBRW} and \cite{HP99}. As we will show below, a possible origin
of this discrepancy might be related to the build-up of a finite
chemical potential, which is usually not included in most 
hydrodynamical calculations to date (see, however,  Ref.~\cite{HS98}). 
   
Recent hadro-chemical analysis~\cite{pbm98,CR99}
of a large body of hadronic heavy-ion
data has shown that the finally observed particle abundances at SpS energies  
are consistent with a common {\em chemical} freezeout at 
temperatures/baryon chemical potentials around $(T,\mu_B)_{ch}=(175,270)$~MeV.  
In the subsequent expansion and cooling the system is still 
strongly interacting with elastic collisions maintaining
thermal equilibrium until the {\em thermal} freezeout. The absence
of pion-number changing (inelastic) reactions (which is well supported
phenomenologially from the inelasticities in $\pi\pi$ scattering)  
then induces a finite pion chemical potential~\cite{Bebie}. 
To incorporate these features, we  here take recourse
to a simple expanding thermal fireball model.  
We start by generating a (non-interacting) resonance gas equation of state 
of hot hadronic matter including the lowest 32 (16) mesonic (baryonic) states.
Imposing entropy and baryon number conservation determines
the temperature dependence of the baryon chemical potential, $\mu_B(T)$.  
The additional requirement of  pion- [kaon-] number conservation 
entails a finite $\mu_\pi(T)$ [$\mu_K(T)\simeq\mu_{\bar K}(T)$], with  
the various baryon chemical potentials
kept in relative equilibrium (with respect to strong interactions), \eg, 
$\mu_\Delta(T)=\mu_N(T)+\mu_\pi(T)$, etc.. With an entropy per baryon
of $s/\varrho_B=26$, our trajectory is found to pass through
$(T,\mu_N)=(175,250)$~MeV, compatible with the experimentally 
deduced point~\cite{pbm98} (here, a small $\mu_\pi\simeq 20$~MeV is needed
to be consistent with the final pion-to-baryon ratio of about 5:1 at 
SpS energies). 
Towards thermal freezeout at $(T,\mu_N)_{fo}\simeq (110-120,415-450)$~MeV, 
$\mu_\pi(T)$  and $\mu_K(T)$ increase approximately linearly 
to around 80 and 110~MeV, respectively.  
Finally we need to introduce a time scale to obtain the volume expansion.
We approximate the latter by a cylindrical geometry as
\beq
V_{FC}^{(2)}(t)= 2 \ (z_0+v_z t +\frac{1}{2} a_z t^2) \ \pi \
(r_0+\frac{1}{2} a_\perp t^2)^2 
\eeq
employing two firecylinders expanding
in the $\pm z$ direction. Guided by hydrodynamical simulations~\cite{HS98}
the primordial longitudinal motion for Pb(158~AGeV)+Au reactions 
is taken to be $v_{z}=0.5c$, and the longitudinal and transverse 
acceleration are fixed to give final velocities $v_{z}(t_{fo})\simeq 0.75c$,
$v_\perp(t_{fo})\simeq 0.55c$ as borne out from experiment (this, in turn, 
requires fireball lifetimes of about $t_{fo}=10-12$~fm/c and implies 
transverse expansion by 3-4~fm, consistent with HBT analyses~\cite{hbt}).  
The parameter $r_0$ denotes the initial nuclear overlap radius, 
\eg, $r_0=4.6$~fm for collisions with impact parameter $b=5$~fm and 
$N_B\simeq 260$ participant baryons. The parameter $z_0$ is equivalent to 
a formation time and fixes the starting point of the trajectory
in the $(T,\mu_N)$ plane. 

With the such specified space-time evolution for $\sim$30\% 
central Pb(158~AGeV)+Au collisions
the dilepton yields from in-medium radiation are obtained as
\bea
\frac{d^2N_{ee}}{d\eta dM} &=&
\frac{\alpha^2}{\pi^3 g_{\rho\pi\pi}^2 M} \int\limits_0^{t_{fo}} 
dt \ V_{FC}^{(1)}(t)
\int \frac{d^3q}{q_0} \ f^\rho(q_0;T(t)) \
\nonumber\\
 & & \times {\cal F}(M,q; \mu_B(t),T(t)) \ Acc(M,\vec q) \ 
\label{dNdM_med}
\eea
with $Acc(M,\vec q)$ accounting for the experimental acceptance cuts (as 
well as an approximate mass resolution) in the '96 CERES data. The backward
firecylinder is placed at a rapidity $y=2.6$ to ensure the correct
charged particle multiplicity as quoted by CERES. The electromagnetic
'transition form factor' ${\cal F}$ 
contains the in-medium $\rho$ spectral function
as described above, where the additional effects   
of the finite meson chemical potentials have been implemented in 
Boltzmann approximation. For initial conditions 
$(T,\varrho_B)_{ini}$=(190~MeV,2.55$\varrho_0$) and a freezeout at
$(T,\varrho_{B})_{fo}$=(115~MeV,0.33$\varrho_0$) the fireball lifetime
amounts to about $t_{fo}$=11~fm/c, see also above
(reducing, \eg, $(T,\varrho_B)_{ini}$ to (180~MeV,1.92$\varrho_0$) reduces
the dilepton signal from the fireball by about 15\%).     
 
\bfig
\bce
\epsfig{file=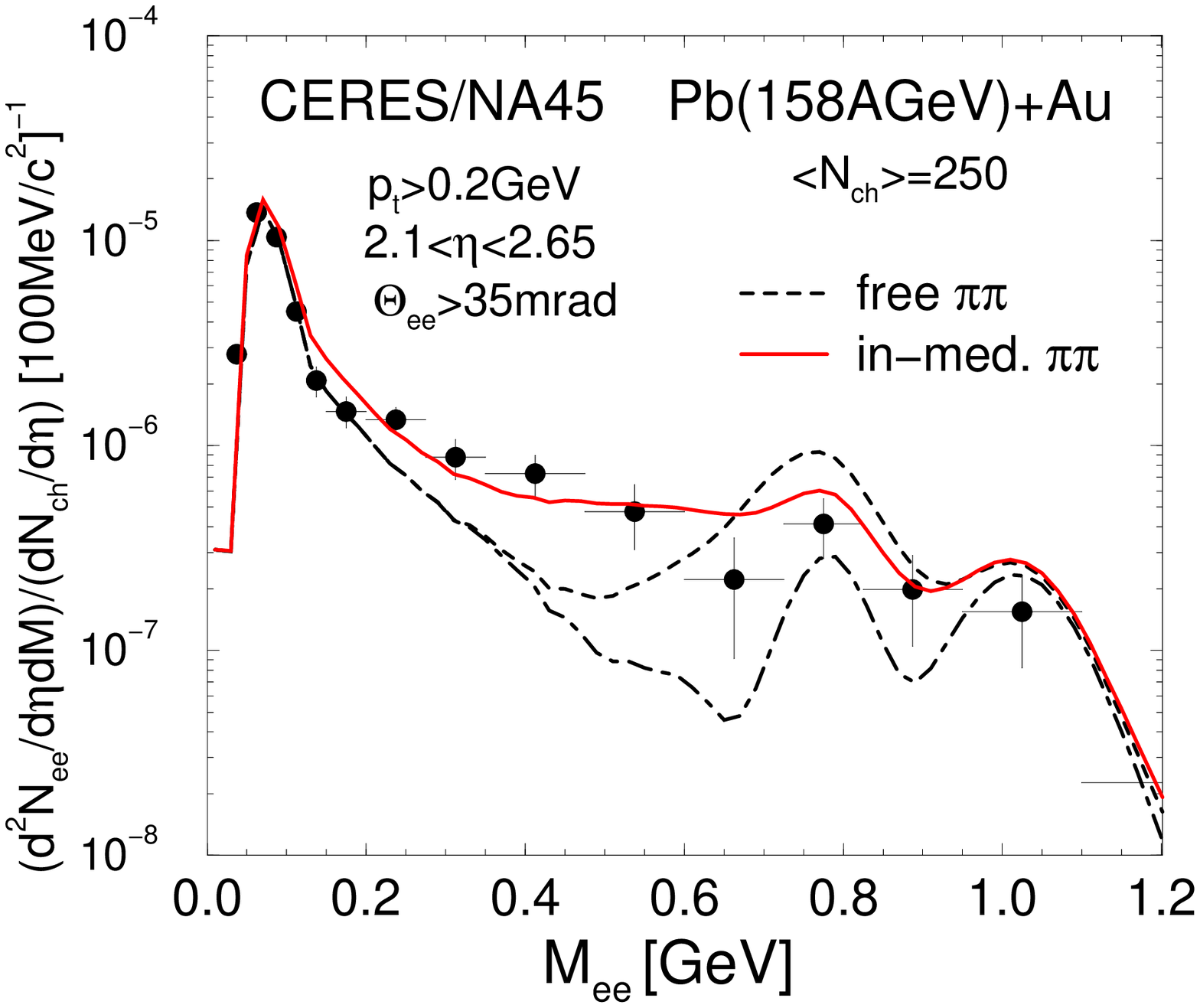,width=8cm}
\ece
\caption{1996 CERES/NA45 inclusive
dielectron invariant mass spectra from 30\% central
Pb(158~AGeV)+Au collisions~\protect\cite{ceres96} compared to the hadronic
cocktail~\protect\cite{cktl} (without $\rho$ decays, dashed-dotted line)
and to the cocktail plus $\pi\pi$ annihilation from the hadronic fireball
using either the free (dashed line) or the in-medium (solid line)
$\rho$ spectral function from Eq.~(\protect\ref{Drho}).}
\label{fig_Mspec}
\efig

In addition to the in-medium radiation described by Eq.~(\ref{dNdM_med})
the finally observed dilepton spectra contain a sizeable contribution
from free hadron decays after freezeout, the so-called 'cocktail'. 
For this we employ the most recent evaluation of the CERES 
collaboration~\cite{cktl},
which is generated from particle abundances based on the chemical
freezeout of Ref.~\cite{pbm98} being consistent with our $(T,\mu_B)$
trajectory. As in our previous works~\cite{RCW} the 
$\rho$-meson decays have been excluded from the cocktail as, owing  
to the short lifetime of the $\rho$,  they are accounted for through 
the in-medium part, Eq.~(\ref{dNdM_med}), in the vicinity of the (idealized) 
'freezeout'.     

The final results for the inclusive invariant mass spectra, shown in 
Fig.~\ref{fig_Mspec}, demonstrate that the use of the in-medium spectral
function leads to reasonable agreement with the '96 CERES data, which 
cannot be described assuming vacuum $\rho$ properties. 
This is in line with the conclusions of our earlier analysis~\cite{RCW},
but we emphasize again that the results presented here are based
on a much improved understanding  of both the microscopic $\rho$ 
spectral function (being constrained by a large body of independent 
data) as well as the space-time evolution dynamics (consistent 
with chemical freezeout analyses, without the use
of an overall normalization factor, etc.). As compared to Ref.~\cite{RCW} 
the smaller $\mu_B$ in the early stages, the finite $\mu_\pi$ in the 
later stages, and the more complete assessment of the meson gas 
effects~\cite{RG99} reduces the dominant role of the baryonic contributions  
to about equal importance as the mesonic ones. 
Another noteworthy point is that the overall dilepton yield exceeds 
what has been found in the  hydrodynamical calculations of, \eg, 
Ref.~\cite{HP99} (where $\mu_\pi=0$ has been used throughout)
by about a factor of $\sim$2.5, thereby  
essentially resolving the above mentioned discrepancy to  
transport calculations~\cite{CEK,LKB}.  

\bfig
\bce
\epsfig{file=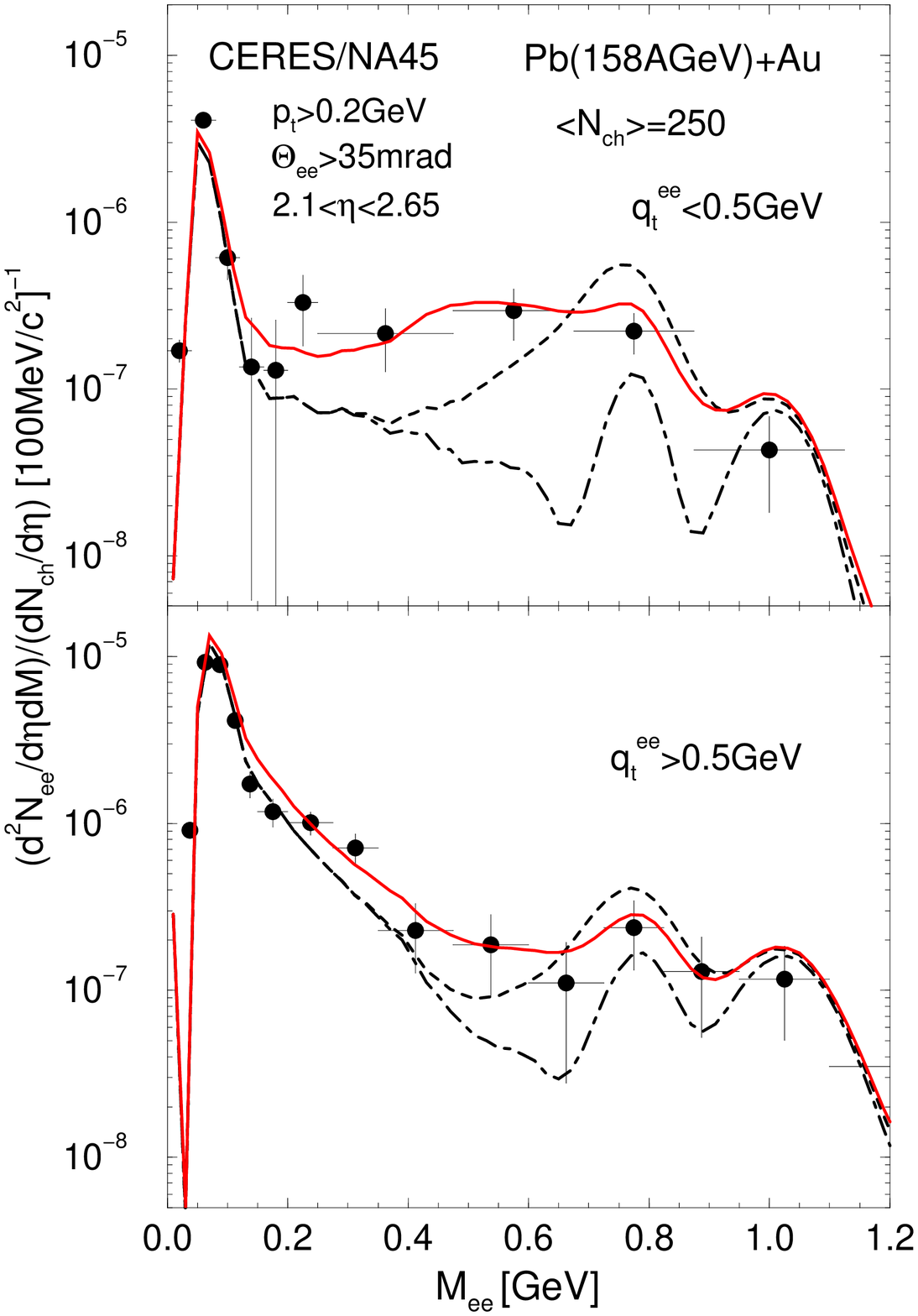,width=8cm}
\ece
\caption{Same as Fig.~\protect\ref{fig_Mspec}, but split into two 
transverse momentum bins~\protect\cite{ceres96}
 below (upper panel) and above (lower panel) $q_t$=0.5~GeV.}
\label{fig_Mspecqt}
\efig

A more detailed insight into the nature of the low-mass enhancement
is revealed by inspecting its dependence on the transverse pair
momentum $q_t$. Fig.~\ref{fig_Mspecqt} shows the invariant 
mass spectrum seperated into two transverse momentum bins below and above
$q_t=0.5$~GeV. Clearly, the excess of the data 
over the free calculation is most pronounced in the
small momentum bin which is properly reflected in our in-medium calculations.
The same behavior is exhibited in another
projection of the data in Fig.~\ref{fig_qtspec}, where transverse
momentum spectra in various invariant mass bins are displayed.
\bfig[!htb]
\bce
\epsfig{file=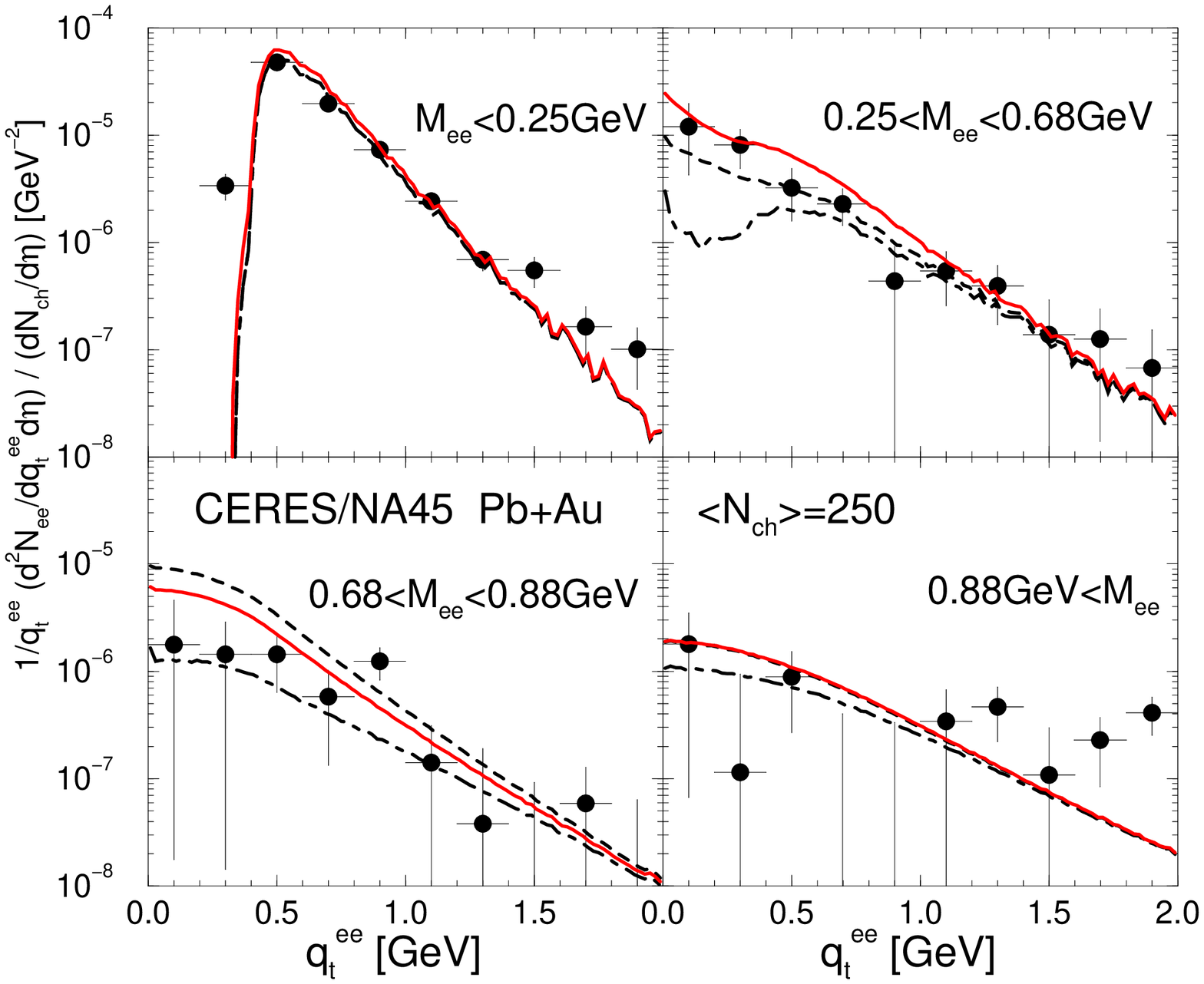,width=8.3cm}
\ece
\caption{1996 CERES/NA45 transverse momentum spectra for $e^+e^-$ pairs
in 30\% central Pb(158~AGeV)+Au collisions~\protect\cite{ceres96}. Line 
identification as in Fig.~\protect\ref{fig_Mspec}.}
\label{fig_qtspec}
\efig

\section{Summary and Discussion}
\label{sec_disc}
Based on realistic hadronic interaction vertices we have shown that 
the in-medium $\rho$ spectral function, evaluated within standard
many-body techniques,  undergoes a strong broadening
in hot and dense matter. At comparable densities the baryonic effects 
prevail over mesonic ones  due to the stronger nature
of the meson-nucleon interactions. Corresponding 
thermal dilepton production rates exhibit a remarkable tendency
to approach the lowest-order perturbative $q\bar q$ 
annihilation with increasing temperature and density. 
This might indicate the emergence of 'quark-hadron duality' towards
the phase boundary down to rather low mass scales of $0.5$~GeV,  
made possible through large imaginary parts resummed in the Dyson
equation of $\rho$ propagator. The medium effects might thus be 
interpreted as a lowering of the 'duality threshold', which in free space
is located around 1.5~GeV. An unambiguous identification of chiral
restoration, however, requires 
the simultaneous treatment of the axial correlator ($a_1$ channel)
which has to merge with the vector correlator which we have focused
on here. Although the perturbative $q\bar q$ rate implies restored chiral
symmetry, one has to keep in mind that (thermal) non-perturbative 
effects below 1~GeV are not yet well under control.  
The argument can be made more rigorous in the 1--1.5~GeV region, 
where the lowest-order in temperature mixing between vector and 
axialvector correlators suffices to establish a full  
degeneracy between them as well as the perturbatively calculated partonic 
result.

We have furthermore demonstrated that our model for the in-medium $\rho$ 
spectral function is in line with current  low-mass
dilepton measurements at the CERN-SpS. For that we have employed  
a simple fireball expansion model consistent with recent hadro-chemical 
analysis and imposing effective pion-number conservation between chemical
and thermal freezeout via finite pion chemical potentials (the 
latter seem to resolve a large part of the discrepancy in the total 
dilepton yield between transport and chemical equilibrium-based hydrodynamic
simulations).  The apparent depletion of the in-medium dilepton yield 
in the  $\rho/\omega$ region, as well as the
enhancement below can be accounted for. This also holds for the low-$q_t$ 
nature of the excess. 
Upcoming high resolution measurements as well as the more baryon-dominated
40~GeV run will put the model predictions under further scrutiny and 
can be expected to advance our understanding of  
chiral symmetry restoration in hot and dense hadronic matter.

\vskip1cm
 
\centerline {\bf ACKNOWLEDGMENTS}
We are grateful for productive conversations with P. Braun-Munzinger,  
G.E. Brown, C. Gale, A. Drees, V. Koch, E.V. Shuryak, H. Sorge, 
H.-J. Specht and J. Stachel.  
We are especially indebted to A. Drees for providing us with the
CERES cocktail. One of us (RR) acknowledges support 
from the Alexander-von-Humboldt foundation as a Feodor-Lynen fellow. 
This work is supported in part by the U.S. Department of Energy 
under Grant No. DE-FG02-88ER40388 and the National Science Foundation
under Grant No. NSF-PHY-98-00978.

\end{document}